\documentclass[aps,pre,amsmath,amssymb,longbibliography,lengthcheck,showpacs,superscriptaddress,floatfix]{revtex4-1}
\usepackage{graphicx,color}

\begin{document}

%%%%%%%%%%%%%%%%%%%%%%%%%%%%%%%%%%%%%%%%%%%%%%%%%%%%%%%%%%%%%%%%%%%%%%%%%%%%%%%%%%%%%%%%%%%%%%%%%%%%%%%%%%%%%%%%%%%%%%%%%%%
\title{Structural and mechanical characteristics of sphere packings near the jamming transition: From fully amorphous to quasi-ordered structures}

%%%%%%%%%%%%%%%%%%%%%%%%%%%%%%%%%%%%%%%%%%%%%%%%%%%%%%%%%%%%%%%%%%%%%%%%%%%%%%%%%%%%%%%%%%%%%%%%%%%%%%%%%%%%%%%%%%%%%%%%%%%
\author{Hideyuki Mizuno}
\affiliation{Graduate School of Arts and Sciences, The University of Tokyo, Tokyo 153-8902, Japan}

\author{Kuniyasu Saitoh}
\affiliation{Department of Physics, Faculty of Science, Kyoto Sangyo University, Kyoto 603-8555, Japan}

\author{Leonardo E.~Silbert}
\affiliation{School of Math, Science, and Engineering, Central New Mexico Community College, Albuquerque, New Mexico 87106, USA}

%%%%%%%%%%%%%%%%%%%%%%%%%%%%%%%%%%%%%%%%%%%%%%%%%%%%%%%%%%%%%%%%%%%%%%%%%%%%%%%%%%%%%%%%%%%%%%%%%%%%%%%%%%%%%%%%%%%%%%%%%%%
\date{\today}

%%%%%%%%%%%%%%%%%%%%%%%%%%%%%%%%%%%%%%%%%%%%%%%%%%%%%%%%%%%%%%%%%%%%%%%%%%%%%%%%%%%%%%%%%%%%%%%%%%%%%%%%%%%%%%%%%%%%%%%%%%%
\begin{abstract}
Mechanically stable sphere packings are generated in three-dimensional space using the discrete element method, which span a wide range in structural order, ranging from fully amorphous to quasi-ordered structures, as characterized by the bond orientational order parameter.
As the packing pressure, $p$, varies from the marginally rigid limit at the jamming transition ($p \approx 0$) to that of more robust systems ($p \gg 0$), the coordination number, $z$, follows a familiar scaling relation with pressure, namely, $\Delta z = z - z_c \sim p^{1/2}$, where $z_c = 2d = 6$~($d=3$ is the spatial dimension).
While it has previously been noted that $\Delta z$ does indeed remain the control parameter for determining the packing properties, here we show how the packing structure plays an influential role on the mechanical properties of the packings.
Specifically, we find that the elastic (bulk $K$ and shear $G$) moduli, generically referred to as $M$, become functions of both $\Delta z$ and the structure, to the extent that $M-M_c \sim \Delta z$.
Here, $M_c$ are values of the elastic moduli at the jamming transition, which depend on the structure of the packings.
In particular, the zero shear modulus, $G_c=0$, is a special feature of fully amorphous packings, whereas more ordered packings take larger, positive values, $G_c > 0$.
\end{abstract}

%%%%%%%%%%%%%%%%%%%%%%%%%%%%%%%%%%%%%%%%%%%%%%%%%%%%%%%%%%%%%%%%%%%%%%%%%%%%%%%%%%%%%%%%%%%%%%%%%%%%%%%%%%%%%%%%%%%%%%%%%%%
\maketitle

%%%%%%%%%%%%%%%%%%%%%%%%%%%%%%%%%%%%%%%%%%%%%%%%%%%%%%%%%%%%%%%%%%%%%%%%%%%%%%%%%%%%%%%%%%%%%%%%%%%%%%%%%%%%%%%%%%%%%%%%%%%
\section{Introduction}
Many previous studies~(e.g., \cite{Ohern_2003,Silbert_2005,Ellenbroek_2006,Silbert_2009}) have established peculiar mechanical and vibrational properties of \textit{disordered} particulate systems close to the jamming transition.
The elastic (bulk $K$ and shear $G$) moduli, generically referred to as $M$, follow power-law scalings with the packing pressure $p$, and in particular, the shear modulus continuously vanishes when approach the transition, as $G \sim p^{1/2}$.
Additionally, the vibrational density of states (vDOS) exhibits a characteristic plateau above the frequency $\omega^\ast$, which goes to zero, following a power-law scaling of $\omega^\ast \sim p^{1/2}$.
These critical behaviors of $M$ and $\omega^\ast$ can be explained by ``isostaticity", where the excess contact number $\Delta z = z -z_c$ ($z_c=2d$ is an isostatic number and $d$ is the spatial dimension) is a central parameter for controlling the material properties~\cite{Wyart_2005,Wyart_2006,Wyart_2010,DeGiuli_2014,Yan_2016}.
Both the shear modulus $G$ and the frequency $\omega^\ast$ are linearly scaled by $\Delta z \sim p^{1/2}$, i.e., $G \sim \Delta z$ and $\omega^\ast \sim \Delta z$.
Interestingly, the same scaling laws were also found in dimer packings~\cite{Schreck_2010,Shiraishi_2019,Shiraishi_2020}.

Similar to disordered systems, even \textit{ordered} particulate systems are shown to exhibit critical scaling laws near the jamming transition~\cite{Silbert_2006,Goodrich_2014,Tong_2015,Charbonneau_2019,Tsekenis_2020,Ikeda_2020}.
A seminal work~\cite{Silbert_2006} systematically modified the structure of the system by introducing ``disorder" and studied the effects of structural modifications on the distributions of the contact number and contact force.
More recently, Goodrich~\textit{et al.}~\cite{Goodrich_2014} demonstrated that although \textit{perfectly} ordered crystals never show any critical behavior, only a small amount of disorder is enough to make the system behave as a highly disordered systems.
In addition, Tong~\textit{et al.}~\cite{Tong_2015} modified the structure by introducing a polydispersity $\eta$ and controlling $\eta$ and established a phase diagram in the packing pressure ($p$) and the polydispersity ($\eta$) plane that identifies three phases, i.e., the crystal, disordered crystal, and amorphous phases.
They demonstrated that even disordered crystals, which maintain an ordered lattice structure, show critical scaling behaviors near the jamming.
Most recently, Tsekenis~\textit{et al.}~\cite{Charbonneau_2019,Tsekenis_2020} showed that such disordered crystals exhibit a power-law scaling in force and gap distributions and a plateau in the vDOS, as do fully amorphous systems.
Finally, using a model of perceptron~\cite{Franz_2015}, Ikeda~\cite{Ikeda_2020} theoretically demonstrated that even weakly disordered crystals show jamming scaling laws.
Therefore, it is now established that even ordered~(but not \textit{perfectly} ordered) systems behave as highly disordered systems near the jamming transition.

\begin{figure}[t]
\centering
\includegraphics[width=0.375\textwidth]{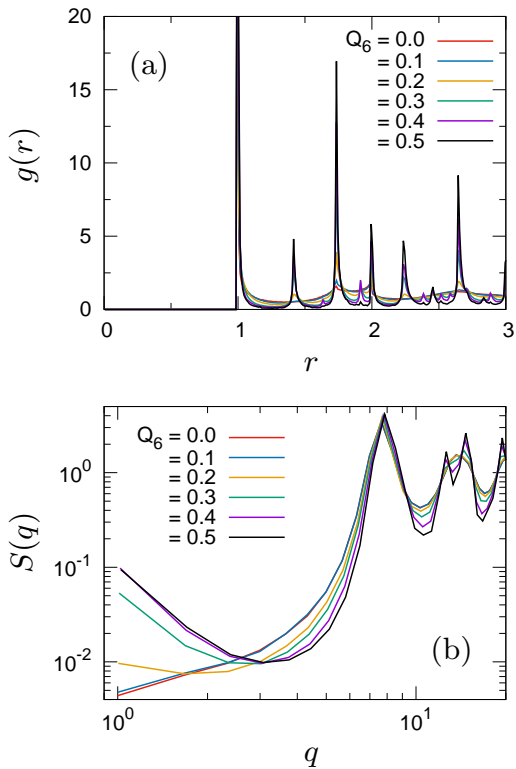}
\vspace*{0mm}
\caption{\label{fig.structure}
Static structures in sphere packings for different values of $Q_6$.
The packing pressure is $p = 4\times 10^{-6}$.
We present the radial distribution function $g(r)$ in (a) and the static structure factor $S(q)$ in (b).
}
\end{figure}

However, in this paper, we will demonstrate that structural properties also play an important role in determining the material properties of the systems.
We analyze jammed particulate systems composed of monodisperse, frictionless, Hookean particles.
We prepare a wide range of structures, ranging from fully amorphous to quasi-ordered structures, which are distinguished by the orientational order parameter, $Q_6 = 0.0$ (disordered) to $0.5$ (ordered)~\cite{Steinhardt_1983}.
For $Q_6 = 0.0$, the system is highly disordered, whereas the crystalline-like, ordered, lattice structure is observed for $Q_6=0.5$ (see Fig.~\ref{fig.structure}).
In this situation, the material properties of the systems generally depend on the packing pressure $p$ as well as the structure $Q_6$.
We observe that the excess contact number $\Delta z$ is always scaled as $\Delta z \sim p^{1/2}$, regardless of the value of $Q_6$.
Our main result is that the elastic moduli, $M= K$ (bulk modulus) and $G$ (shear modulus), are described as functions of $\Delta z$ and $Q_6$:
\begin{equation}
\begin{aligned}
& M(\Delta z, Q_6) = M_c(Q_6) + \alpha_M \Delta z,  \label{equationofKG}
\end{aligned}
\end{equation}
where $M_c= K_c, G_c$ are critical values at the jamming transition, and $\alpha_M = \alpha_K, \alpha_G$ are constants.
Therefore, $M - M_c \sim \Delta z$ is controlled by $\Delta z$ only and shows the same critical scaling regardless of the structure $Q_6$.
However, structural effects appear in $M_c$ at the transition, and a shear modulus that is equal to zero $G_c=0$ is a special feature of disordered packings~\cite{Zaccone_2011,Zaccone2_2011,Mizuno3_2016,Cui_2019}.
For quasi-ordered packings, the shear modulus becomes finite and positive, $G_c > 0$.
We will also show that this difference in $G_c$ is reflected in the vibrational states of the systems.

%%%%%%%%%%%%%%%%%%%%%%%%%%%%%%%%%%%%%%%%%%%%%%%%%%%%%%%%%%%%%%%%%%%%%%%%%%%%%%%%%%%%%%%%
%%%%% Fig. 1 %%%%%
%%%%%%%%%%%%%%%%%%%%%%%%%%%%%%%%%%%%%%%%%%%%%%%%%%%%%%%%%%%%%%%%%%%%%%%%%%%%%%%%%%%%%%%%

%%%%%%%%%%%%%%%%%%%%%%%%%%%%%%%%%%%%%%%%%%%%%%%%%%%%%%%%%%%%%%%%%%%%%%%%%%%%%%%%%%%%%%%%
%%%%% Fig. 2 %%%%%
\begin{figure}[t]
\centering
\includegraphics[width=0.375\textwidth]{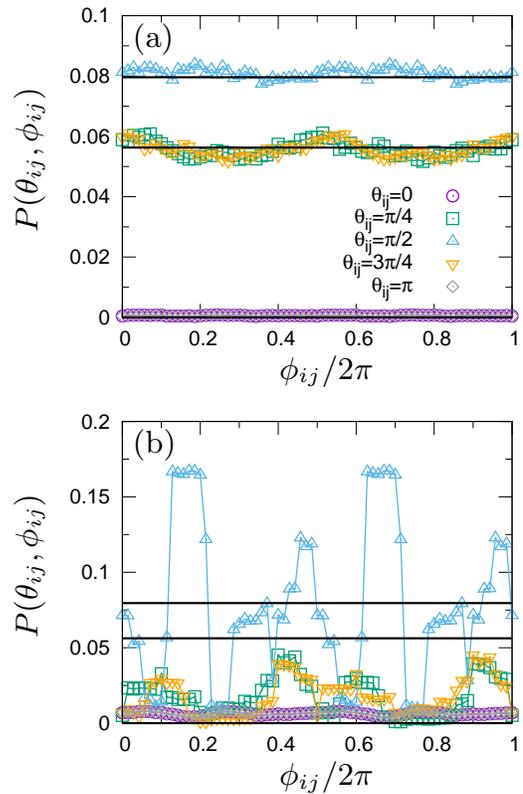}
\vspace*{0mm}
\caption{\label{fig.bond}
Probability distribution function $P(\theta_{ij}, \phi_{ij})$ of the orientation angles of the unit bond vector $\mathbf{n}_{ij} = (\cos\phi_{ij}\sin\theta_{ij},\sin\phi_{ij}\sin\theta_{ij},\cos\theta_{ij})$ for $Q_6 = 0.0$ in (a) and $Q_6 = 0.5$ in (b).
Note $0 \le \phi_{ij} < 2\pi$ and $0 \le \theta_{ij} \le \pi$.
The packing pressure is $p = 4 \times 10^{-6}$.
The solid lines demonstrate the random, isotropic distribution, which coincides well with $P(\theta_{ij}, \phi_{ij})$ for the $Q = 0.0$ (disordered) case.
}
\end{figure}
%%%%%%%%%%%%%%%%%%%%%%%%%%%%%%%%%%%%%%%%%%%%%%%%%%%%%%%%%%%%%%%%%%%%%%%%%%%%%%%%%%%%%%%%

%%%%%%%%%%%%%%%%%%%%%%%%%%%%%%%%%%%%%%%%%%%%%%%%%%%%%%%%%%%%%%%%%%%%%%%%%%%%%%%%%%%%%%%%
%%%%% Fig. 3 %%%%%
\begin{figure*}[t]
\centering
\includegraphics[width=0.75\textwidth]{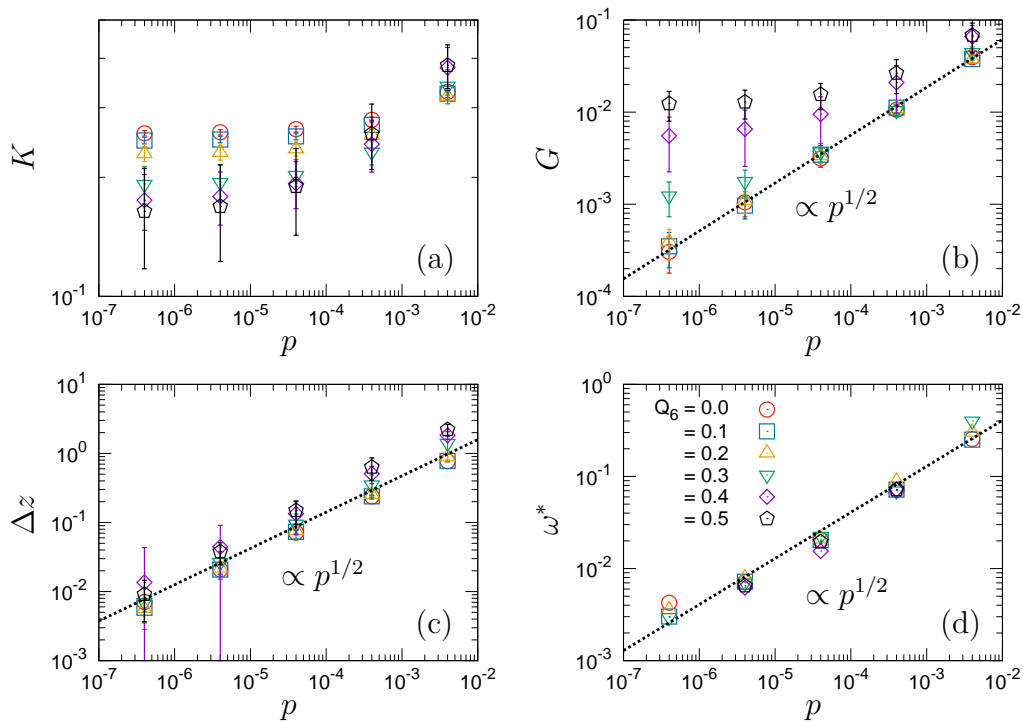}
\vspace*{0mm}
\caption{\label{fig.pdependence}
Dependences on the packing pressure $p$ of the quantities for different $Q_6$ values.
We plot the (a) bulk modulus $K$, (b) shear modulus $G$, (c) excess contact number $\Delta z=z-z_c$, and (d) characteristic frequency in vDOS, $\omega^\ast$, as functions of $p$.
The lines represent power-law scalings with respect to $p$.
The error bars in (a)-(c) are calculated from $100$ configuration realizations.
}
\end{figure*}
%%%%%%%%%%%%%%%%%%%%%%%%%%%%%%%%%%%%%%%%%%%%%%%%%%%%%%%%%%%%%%%%%%%%%%%%%%%%%%%%%%%%%%%%

%%%%%%%%%%%%%%%%%%%%%%%%%%%%%%%%%%%%%%%%%%%%%%%%%%%%%%%%%%%%%%%%%%%%%%%%%%%%%%%%%%%%%%%%
%%%%% Fig. 4 %%%%%
\begin{figure*}[t]
\centering
\includegraphics[width=0.75\textwidth]{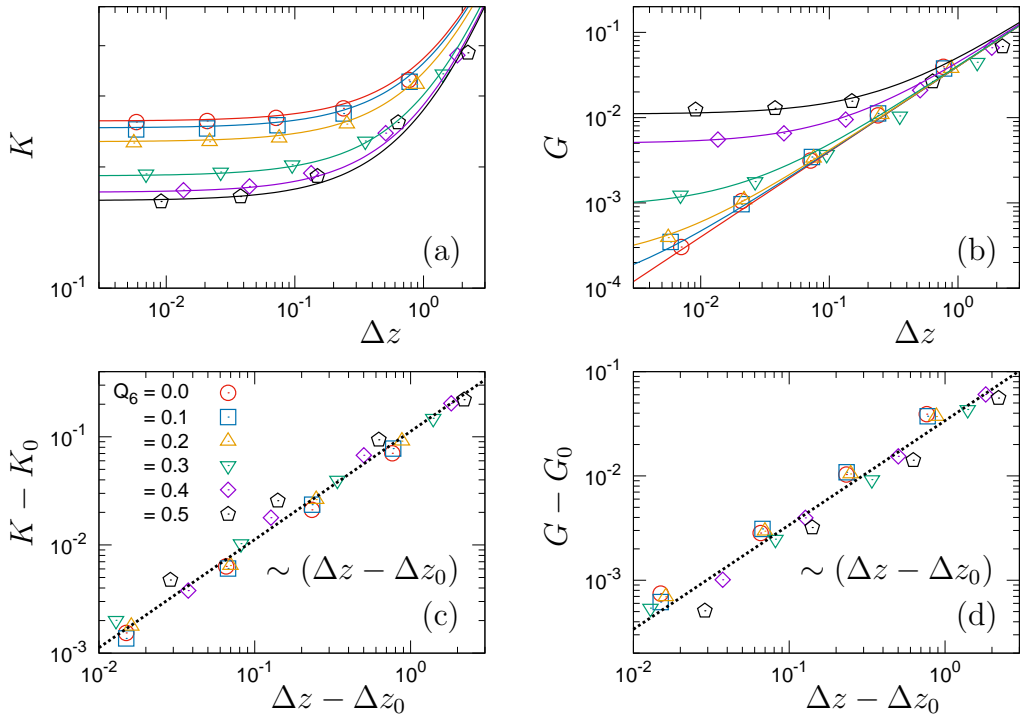}
\vspace*{0mm}
\caption{\label{fig.zdependence}
Dependences on the excess contact number $\Delta z$ of the elastic moduli $K$ and $G$ for different $Q_6$ values.
We plot the (a) bulk modulus $K$ and (b) shear modulus $G$ as functions of $\Delta z$.
In (c) and (d), we plot $K-K_0$ and $G-G_0$ as functions of $\Delta z-\Delta z_0$, where the subscript ``$0$" denotes values at the lowest pressure $p=4\times 10^{-7}$.
The data and symbols are the same as those in Fig.~\ref{fig.pdependence}.
The lines represent $K = K_c + \alpha_K \Delta z$ in (a), $G = G_c + \alpha_G \Delta z$ in (b), $K-K_0 = \alpha_K (\Delta z -\Delta z_0)$ in (c) and $G-G_0 = \alpha_G (\Delta z -\Delta z_0)$ in (d).
Here, $\alpha_K \simeq 0.11$ and $\alpha_G \simeq 0.04$, and the critical values of $K_c$ and $G_c$ are plotted as functions of $Q_6$ in Fig.~\ref{fig.critical}.
}
\end{figure*}
%%%%%%%%%%%%%%%%%%%%%%%%%%%%%%%%%%%%%%%%%%%%%%%%%%%%%%%%%%%%%%%%%%%%%%%%%%%%%%%%%%%%%%%%

%%%%%%%%%%%%%%%%%%%%%%%%%%%%%%%%%%%%%%%%%%%%%%%%%%%%%%%%%%%%%%%%%%%%%%%%%%%%%%%%%%%%%%%%
%%%%% Fig. 5 %%%%%
\begin{figure}[t]
\centering
\includegraphics[width=0.375\textwidth]{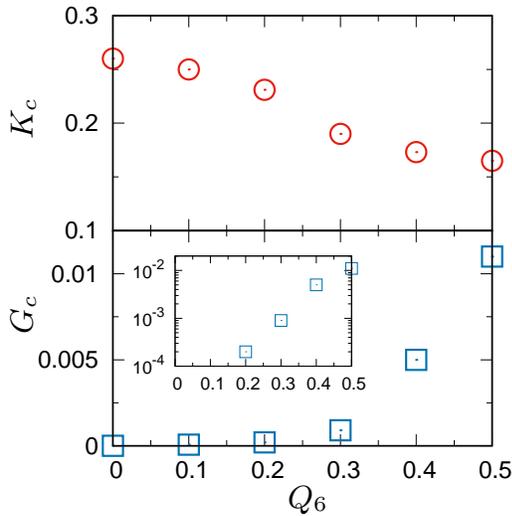}
\vspace*{0mm}
\caption{\label{fig.critical}
Dependences on the structure $Q_6$ of the critical values of the elastic moduli $K_c$ (upper panel) and $G_c$ (lower panel).
The inset to the lower panel plots $G_c$ on the log scale.
}
\end{figure}
%%%%%%%%%%%%%%%%%%%%%%%%%%%%%%%%%%%%%%%%%%%%%%%%%%%%%%%%%%%%%%%%%%%%%%%%%%%%%%%%%%%%%%%%

%%%%%%%%%%%%%%%%%%%%%%%%%%%%%%%%%%%%%%%%%%%%%%%%%%%%%%%%%%%%%%%%%%%%%%%%%%%%%%%%%%%%%%%%
%%%%% Fig. 6 %%%%%
\begin{figure*}[t]
\centering
\includegraphics[width=0.75\textwidth]{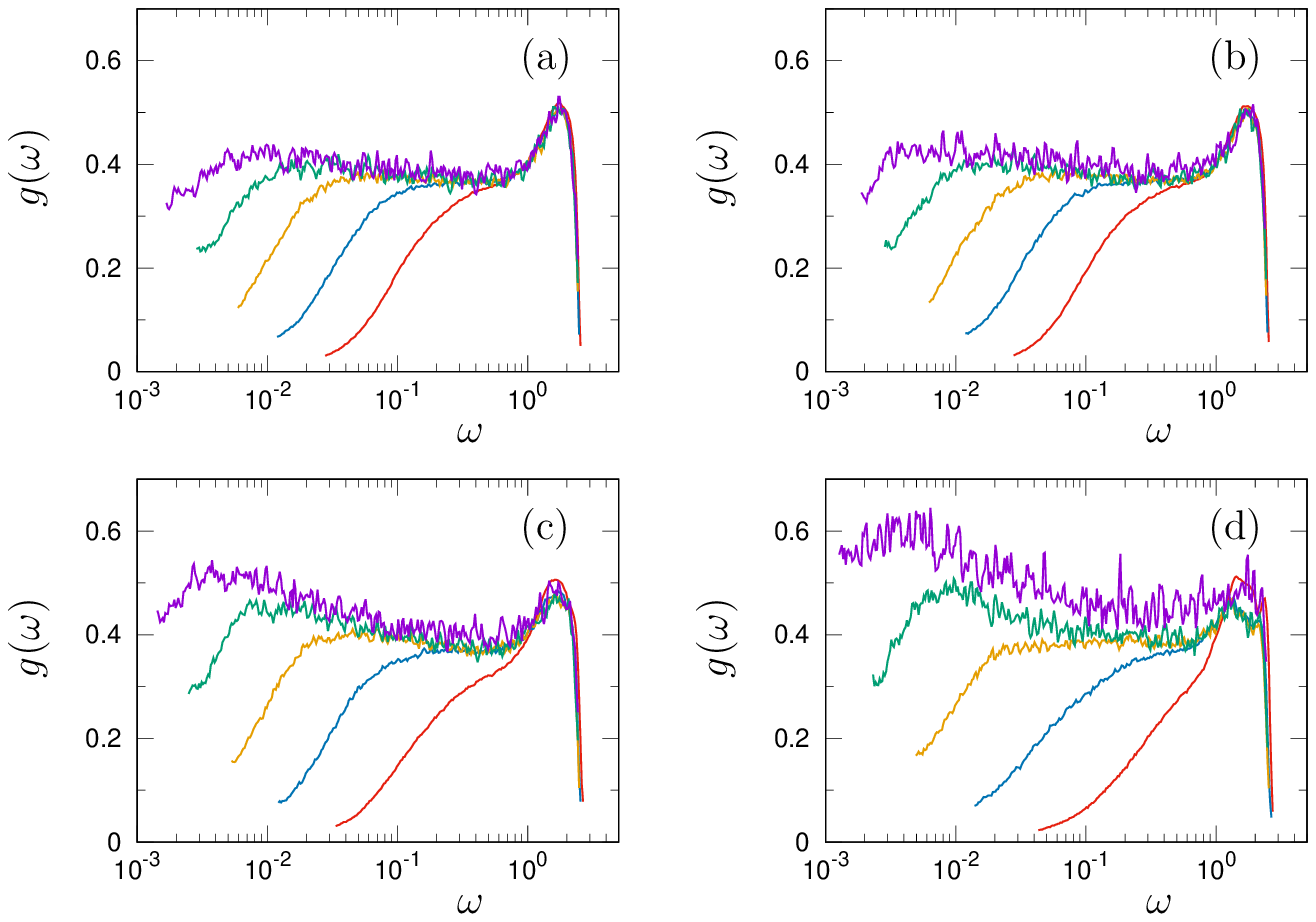}
\vspace*{0mm}
\caption{\label{fig.vibration}
The vDOS for different $Q_6$ values.
We plot $g(\omega)$ as a function of the frequency $\omega$ for $Q_6= 0.0$ in (a), $0.1$ in (b), $0.3$ in (c), and $0.5$ in (d).
Lines of different colors represent different packing pressures, $p = 4\times 10^{-3}$ (red), $4\times 10^{-4}$ (blue), $4\times 10^{-5}$ (orange), $4\times 10^{-6}$ (green), and $4\times 10^{-7}$ (purple), from right to left.
}
\end{figure*}
%%%%%%%%%%%%%%%%%%%%%%%%%%%%%%%%%%%%%%%%%%%%%%%%%%%%%%%%%%%%%%%%%%%%%%%%%%%%%%%%%%%%%%%%

%%%%%%%%%%%%%%%%%%%%%%%%%%%%%%%%%%%%%%%%%%%%%%%%%%%%%%%%%%%%%%%%%%%%%%%%%%%%%%%%%%%%%%%%%%%%%%%%%%%%%%%%%%%%%%%%%%%%%%%%%%%
\section{Simulation model}
%
%%%%%%%%%%%%%%%%%%%%%%%%%%%%%%%%%%%%%%%%%%%%%%%%%%%%%%%%%%%%%%%%%%%%%%%%%%%%%%%%%%%%%%%%%%%%%%%%%%%%%%%%%%%%%%%%%%%%%%%%%%%
\subsection{System preparation}
Our system is composed of $N=1,000$, monodisperse, frictionless particles with a mass $m$ and diameter $\sigma$ in three ($d=3$) dimensional space under periodic boundary conditions.
The particles interact via a finite-range, purely repulsive, harmonic potential, which has been employed in many previous simulations~(e.g., ~\cite{Ohern_2003,Silbert_2005,Ellenbroek_2006,Silbert_2009}):
\begin{equation} \label{interaction}
\phi(r) =
\left\{ \begin{aligned}
& \frac{ \textrm{k} }{2} \left(\sigma-r \right)^2 & (r < \sigma), \\
& 0 & (r \ge \sigma),
\end{aligned} \right. \\
\end{equation}
where $r$ is the distance between two particles, and $\textrm{k}$ parameterizes the particle stiffness and sets an energy scale through $\epsilon = \textrm{k}\sigma^{2}$.
Throughout this paper, we use $\sigma$, $m$, and $\tau=(m/\textrm{k})^{1/2}$ as units of length, mass, and time, respectively, i.e., we set $\sigma=m=\textrm{k} = \tau =\epsilon=1$.

We prepare sphere packings with different structures, which are characterized by the orientational order parameter, $Q_6 = 0.0$ (disordered) to $0.5$ (ordered)~\cite{Steinhardt_1983}.
Here, we use a thermal decompression protocol, which has been employed as ``Protocol 1" in Ref.~\cite{Schreck2_2011}.
Briefly, we prepared equilibrated liquid configurations at a temperature of $T=10^{-3}$ and then quenched them to a very low temperature, $T=10^{-16}$, by changing the cooling rate.
The slower rate creates more ordered configurations (the larger $Q_6$), whereas the faster rate leads to disordered packings (the smaller $Q_6$).

These packings are then put into the ``packing finder" (compression/decompression routine)~\cite{Ohern_2003} and brought to the jamming transition point (where the pressure is $p \approx 10^{-8}$).
Finally, we generate the final configurations at several different packing pressures $p$ by compressing the systems from the jamming transition.
Note that we always remove the rattler particles that have fewer than $d=3$ contacting neighbors.
A total of $100$ configuration realizations are prepared at each $p$ and each $Q_6$, and the values of the physical quantities presented below (e.g., the elastic moduli $M$) are obtained by taking the average of these $100$ realizations.

%%%%%%%%%%%%%%%%%%%%%%%%%%%%%%%%%%%%%%%%%%%%%%%%%%%%%%%%%%%%%%%%%%%%%%%%%%%%%%%%%%%%%%%%%%%%%%%%%%%%%%%%%%%%%%%%%%%%%%%%%%%
\subsection{Structural characteristics}
Figure~\ref{fig.structure} presents the radial distribution function $g(r)$~in (a) and the static structure factor $S(q)$~in (b) for different $Q_6$ values and a pressure of $p = 4\times 10^{-6}$.
In the case of $Q_6 = 0.0$, we see a highly disordered structure~\cite{Silbert2_2006}.
However, as $Q_6$ increases toward $0.5$, the system becomes a more ordered state.
Indeed, we can clearly observe sharp peaks in $g(r)$, which is a feature of the crystalline-like, ordered, lattice structure.
Additionally, $S(q)$ shows clear enhancement of the long-range spatial correlation at small wavenumbers $q$.

In addition, Figure~\ref{fig.bond} shows the probability distribution of the unit bond vector $\mathbf{n}_{ij}$ of connected particles $i$ and $j$.
Here, we define the bond vector as $\mathbf{n}_{ij} = (n_{ij}^x,n_{ij}^y,n_{ij}^z) = (\cos\phi_{ij}\sin\theta_{ij},\sin\phi_{ij}\sin\theta_{ij},\cos\theta_{ij})$ and show the joint probability distribution $P(\theta_{ij}, \phi_{ij})$~(see Ref.~\cite{Mizuno3_2016} for details).
For the case of $Q_6=0.0$ in (a), we clearly observe a random, isotropic distribution, $P(\theta_{ij}, \phi_{ij}) = (1/2\pi)(\sin \theta_{ij}/2)$~\cite{Zaccone_2011,Zaccone2_2011,Mizuno3_2016,Cui_2019}.
In contrast, for the ordered case of $Q=0.5$ in (b), the distribution is completely different from this isotropic distribution.
The pronounced values in $P(\theta_{ij}, \phi_{ij})$ imply ordered structures, which is consistent with the indication of $Q_6$.

%%%%%%%%%%%%%%%%%%%%%%%%%%%%%%%%%%%%%%%%%%%%%%%%%%%%%%%%%%%%%%%%%%%%%%%%%%%%%%%%%%%%%%%%%%%%%%%%%%%%%%%%%%%%%%%%%%%%%%%%%%%
\section{Results}
In the present work, we study the mechanical and vibrational properties of disordered ($Q_6=0.0$) to ordered ($Q_6=0.5$) systems and clarify their dependences on the $Q_6$ value.
The elastic moduli, the bulk $K$ and shear $G$ moduli, are calculated by using the harmonic formulation~\cite{Lemaitre_2006}.
In this formulation, we can calculate elastic moduli without applying any explicit deformation field~(details are found in Ref.~\cite{Mizuno3_2016}).
Additionally, we diagonalize the Hessian matrix to obtain vibrational eigenmodes and calculate the vDOS, $g(\omega)$ and its characteristic frequency $\omega^\ast$.
Figure~\ref{fig.pdependence} plots the elastic moduli, $K,G$, excess contact number, $\Delta z = z - z_c$, and frequency, $\omega^\ast$, as functions of the packing pressure $p$, for different $Q_6$ values.

%%%%%%%%%%%%%%%%%%%%%%%%%%%%%%%%%%%%%%%%%%%%%%%%%%%%%%%%%%%%%%%%%%%%%%%%%%%%%%%%%%%%%%%%%%%%%%%%%%%%%%%%%%%%%%%%%%%%%%%%%%%
\subsection{Excess contact number $\boldsymbol{\Delta z}$}
We first look at the contact number and find that it takes the value of $z_c=2d(1-N^{-1}) \approx 6.0$ at the transition, regardless of the structural properties ($Q_6$).
Then, the excess contact number, $\Delta z = z -z_c$, follows the same power-law scaling trend, $\Delta z \sim p^{1/2}$, for all of the studied $Q_6$ values (see Fig.~\ref{fig.pdependence}(c)).
We note that \textit{perfectly} ordered crystals can\textit{not} show such critical behavior.
However, quasi-ordered systems, which are not perfectly ordered but exhibit crystalline lattice structures, as shown in Fig.~\ref{fig.structure}, can show the scaling law of $\Delta z \sim p^{1/2}$.
This observation is consistent with previous simulation results~\cite{Goodrich_2014,Tong_2015}.
Ref.~\cite{Goodrich_2014} found that only a small amount of disorder makes the system behave as a highly disordered system to exhibit the jamming scaling law.
Additionally, Ref.~\cite{Tong_2015} demonstrated that for the case in which polydispersity produces spatial fluctuations in the distribution of the contact number, even the system with ordered lattice structure shows critical scaling.

%%%%%%%%%%%%%%%%%%%%%%%%%%%%%%%%%%%%%%%%%%%%%%%%%%%%%%%%%%%%%%%%%%%%%%%%%%%%%%%%%%%%%%%%%%%%%%%%%%%%%%%%%%%%%%%%%%%%%%%%%%%
\subsection{Elastic moduli $\boldsymbol{M = K, G}$}
In Figs.~\ref{fig.pdependence}(a) and (b), we clearly observe that the elastic moduli $K$ and $G$ depend on the structural properties $Q_6$.
Particularly, when approaching the jamming transition as $p \rightarrow 0$, the shear modulus $G$ vanishes continuously, following $G\sim p^{1/2}$, in a fully amorphous state of $Q_6=0.0$, whereas it converges to a finite value for the ordered cases with $Q_6 > 0$.
We also plot $K$ and $G$ as functions of $\Delta z$ instead of $p$, in Figs.~\ref{fig.zdependence}(a) and (b) (symbols).
As validated below, we can describe $K$ and $G$ as functions of $\Delta z$ and $Q_6$~[Eq.~(\ref{equationofKG})]:
\begin{equation}
\begin{aligned}
K(\Delta z, Q_6) &= K_c(Q_6) + \alpha_K \Delta z, \\
G(\Delta z, Q_6) &= G_c(Q_6) + \alpha_G \Delta z, \label{equationofKG2}
\end{aligned}
\end{equation}
where $\alpha_K \simeq 0.11$ and $\alpha_G \simeq 0.04$ are constants.

To validate Eq.~(\ref{equationofKG2}), Fig.~\ref{fig.zdependence} plots $K-K_0$ in (c) and $G-G_0$ in (d) as functions of $\Delta z-\Delta z_0$, where the subscript ``$0$" denotes values at the lowest pressure, $p=4 \times 10^{-7}$.
Both $K-K_0$ and $G-G_0$ conveniently collapse on a single curve as a function of $\Delta z-\Delta z_0$ for different $Q_6$ values:
\begin{equation}
\begin{aligned}
K-K_0 &= \alpha_K (\Delta z -\Delta z_0),\\
G-G_0 &= \alpha_G (\Delta z -\Delta z_0),
\end{aligned}
\end{equation}
which determine the values of $\alpha_K \simeq 0.11$ and $\alpha_G \simeq 0.04$.
In Figs.~\ref{fig.zdependence}(a) and (b) (lines), we also plot Eq.~(\ref{equationofKG2}) to the numerical data of $K$ and $G$ by using fixed values of $\alpha_K \simeq 0.11$ and $\alpha_G \simeq 0.04$ and adjusting the values of $K_c$ and $G_c$.
Eq.~(\ref{equationofKG2})~(lines) fits well to the numerical data~(symbols) for all the $Q_6$ cases, where $K_c$ and $G_c$ are determined as functions of $Q_6$, as plotted in Fig.~\ref{fig.critical}.
These results validate Eq.~(\ref{equationofKG2}) for the elastic moduli $K$ and $G$.

Eq.~(\ref{equationofKG2}) separates the dependences of $K,G$ on the excess contact number $\Delta z$ from those on the structure $Q_6$.
Interestingly, the scaling behaviors of $K-K_c$ and $G-G_c$ are both only controlled by $\Delta z$, regardless of $Q_6$.
This result indicates that the isostaticity~\cite{Wyart_2005,Wyart_2006,Wyart_2010,DeGiuli_2014,Yan_2016} controls the mechanical properties near the jamming transition, regardless of whether the systems are disordered or ordered systems.
However, structural effects emerge for critical values of $K_c$ and $G_c$ at the transition.

Figure~\ref{fig.critical} plots $K_c$ and $G_c$ as functions of $Q_6$.
The bulk modulus $K_c$ decreases as $Q_6$ increases from $Q_6=0.0$ (disordered) to $0.5$ (ordered).
In contrast, the shear modulus $G_c$ increases with increasing $Q_6$.
These tendencies of a more ordered system with a smaller bulk modulus and larger shear modulus are also observed in atomic glasses~\cite{Mizuno2_2013}.
The zero critical value of $G_c = 0$ is a particular feature of fully amorphous packings ($Q_6=0.0$), which is based on the random and isotropic distribution of the bond vectors between the particles in contact~(see Figs.~\ref{fig.bond}(a)).
A detailed discussion on this point is given in our previous work~\cite{Mizuno3_2016}.
In contrast, for quasi-ordered packings, the bond distribution is neither random nor isotropic~(see Figs.~\ref{fig.bond}(b)), which produces the finite value of $G_c>0$.

%%%%%%%%%%%%%%%%%%%%%%%%%%%%%%%%%%%%%%%%%%%%%%%%%%%%%%%%%%%%%%%%%%%%%%%%%%%%%%%%%%%%%%%%%%%%%%%%%%%%%%%%%%%%%%%%%%%%%%%%%%%
\subsection{Vibrational eigenmodes}
Next, the vDOSs are studied for different structures $Q_6$.
Figure~\ref{fig.vibration} shows the $g(\omega)$ for different values of $Q_6 = 0.0$ to $0.5$.
In previous simulations~\cite{Ohern_2003,Silbert_2005}, the vDOS has been studied in the case of $Q_6 = 0.0$~(disordered packings).
As shown in Fig.~\ref{fig.vibration}(a), $g(\omega)$ shows the characteristic plateau, where the vibrational eigenmodes show floppy-like motions~\cite{Silbert_2009}.
The onset frequency of the plateau, $\omega^\ast$, is controlled by the excess contact number $\Delta z$~\cite{Wyart_2005,Wyart_2006,Wyart_2010,DeGiuli_2014,Yan_2016}.
Upon approaching the transition, $\omega^\ast$ vanishes, following the power-law scaling of $\omega^\ast \sim \Delta z$.

Here, we can recognize the plateau even in ordered packings up to $Q_6=0.5$, as shown in Figs.~\ref{fig.vibration}(b)-(d).
Most of the recent simulations also showed a plateau in polydisperse crystalline systems~\cite{Charbonneau_2019,Tsekenis_2020}.
We note that the plateau in $g(\omega)$ is enhanced for order packings, as discussed below, but the characteristic frequency, $\omega^\ast$, can still be defined for all the cases of $Q_6=0.0$ to $0.5$.
Figure~\ref{fig.pdependence}(d) plots $\omega^\ast$ as a function of $p$ and demonstrates $\omega^\ast \sim p^{1/2}$ and thus $\omega^\ast \sim \Delta z$ for all $Q_6$ cases.

One noteworthy point is that the plateau in $g(\omega)$ is enhanced as the system becomes more ordered with larger $Q_6$ values.
As demonstrated in Figs.~\ref{fig.pdependence} to~\ref{fig.critical}, the shear modulus $G_c$ becomes finite, \textit{not} vanishing, in ordered packings.
This finite shear modulus excites some amount of transverse acoustic modes at low frequencies, which enhances the plateau value of $g(\omega)$.
Therefore, for ordered packings, acoustic modes controlled by the shear modulus $G$ add to the floppy-like modes controlled by the excess contact number $\Delta z$, whereas for disordered packings, the acoustic modes vanish and the floppy-like modes are dominant.

%%%%%%%%%%%%%%%%%%%%%%%%%%%%%%%%%%%%%%%%%%%%%%%%%%%%%%%%%%%%%%%%%%%%%%%%%%%%%%%%%%%%%%%%%%%%%%%%%%%%%%%%%%%%%%%%%%%%%%%%%%%
\section{Conclusion}
In summary, we have studied particulate systems near the jamming transition by varying their structural properties from fully amorphous to quasi-ordered structures.
We found that ``excess" elastic moduli, $\Delta M = M-M_c$~($M=K, G$), follow the scaling law of $\Delta M \sim \Delta z$, regardless of whether there are ordered or disordered structures.
However, the critical values of $M_c$ at the transition depend on the structure.
As the system becomes more ordered, the bulk modulus $K_c$ decreases while the shear modulus $G_c$ increases.
In particular, the zero shear modulus $G_c=0$ is the nature of fully amorphous packings, while ordered packings have a positive shear modulus of $G_c > 0$.
A characteristic plateau in the vDOS and the onset frequency following $\omega^\ast \sim \Delta z$ are observed, which are again common between disordered and ordered systems.
However, for ordered packings, the finite shear modulus $G_c > 0$ induces transverse acoustic modes which add to the floppy-like modes controlled by $\Delta z$ and enhance the plateau in the vDOS.

Our results demonstrate that the scaling laws of the mechanical and vibrational properties, which are controlled by $\Delta z$, are independent of the structural properties.
This is consistent with the theoretical predictions of Refs.~\cite{Wyart_2005,Wyart_2006,Wyart_2010,DeGiuli_2014,Yan_2016}, which do not assume a specific structure type.
However, what we found here is that critical values of $K_c$ and $G_c$ are controlled by the structural properties.
The shear modulus only vanishes at the transition for disordered packing but not for ordered packings.

The present work and previous studies~\cite{Silbert_2006,Goodrich_2014,Tong_2015,Charbonneau_2019,Tsekenis_2020,Ikeda_2020} have established that quasi-ordered systems can behave as highly disordered systems.
It would be interesting to investigate how quasi-ordered systems share the material properties of disordered systems.
For example, recent studies~\cite{Lerner_2016,Mizuno_2017,Wang_2019,Ikeda2_2019} unveiled the existence of localized vibrational modes in disordered systems and their vDOS following $g(\omega) \sim \omega^4$.
In particular, it was found that the localized modes are controlled by the excess contact number near the jamming transition~\cite{Wyart_2005,Wyart_2006,Yan_2016,Shimada2_2018}.
Another anomalous property could be the elastic response~\cite{Leonforte_2005,Lerner_2014,Karimi_2015}, sound attenuation~\cite{Monaco2_2009,Beltukov_2016,Mizuno_2018,Moriel_2019,Wang2_2019,Saitoh_2019}, and anharmonic~(nonlinear) properties, including thermal activation~\cite{Xu_2010,Mizuno_2020,Mizuno2_2020}, contact change~\cite{Schreck_2011,Deen_2014,Tuckman_2020}, and plastic events~\cite{Maloney2_2006,Manning_2011,Dasgupta2012,Gartner_2016,Morse_2020}.
We may expect that quasi-ordered systems share many of these properties and phenomena, which could be addressed in the future.

%%%%%%%%%%%%%%%%%%%%%%%%%%%%%%%%%%%%%%%%%%%%%%%%%%%%%%%%%%%%%%%%%%%%%%%%%%%%%%%%%%%%%%%%%%%%%%%%%%%%%%%%%%%%%%%%%%%%%%%%%%%
\begin{acknowledgments}
This work was supported by JSPS KAKENHI Grant Numbers 18K13464, 19K14670, and 20H01868.
\end{acknowledgments}

%%%%%%%%%%%%%%%%%%%%%%%%%%%%%%%%%%%%%%%%%%%%%%%%%%%%%%%%%%%%%%%%%%%%%%%%%%%%%%%%%%%%%%%%%%%%%%%%%%%%%%%%%%%%%%%%%%%%%%%%%%%
\bibliographystyle{apsrev4-1}
\bibliography{manuscript}

\end{document}